\documentclass[aps,prl,oneside,twocolumn]{revtex4-1}  
\usepackage[draft]{hyperref} 
\usepackage[usenames]{color}
\usepackage{graphics}
\usepackage{amsmath}

\begin{document}
\title{Imaging studies of photodamage and self healing in Disperse Orange 11 dye-doped PMMA}
\author{Benjamin Anderson, Shiva K. Ramini and  Mark G. Kuzyk}
\address{Department of Physics and Astronomy, Washington State University,
Pullman, WA 99164-2814}
\date{\today}

\begin{abstract}

We report on optical imaging studies of self healing after laser-induced photodamage in disperse orange 11 dye doped into PMMA polymer.  In particular, the high spacial image of the damage track made by a line focus pump laser allows the recovery rates to be measured as a function of burn dose using the relationship between transverse distance and pump intensity profile.  The time evolution of the damaged population results in an intensity-independent time constant of $\tau=490.2 \pm 23$ mins, in agreement with independent measurements of the time evolution of amplified spontaneous emission.  Also observed is a damage threshold above which the material does not fully recover.

\vspace{1em}

\end{abstract}

\maketitle

\vspace{1em}
\definecolor{Red}{rgb}{1.0,0,0}

\section{Introduction}

Photodegradation is a processes by which light damages a material, from sunlight fading colors on upholstery in a car, to catastrophic damage to a material in an optical device caused by a laser.  Over the past two decades, research has focused on understanding laser damage in order to find ways to make materials more resistant to such damage,\cite{Exarh98.01, Wood03.01} and to extend the life of laser components or waveguide polymeric devices.\cite{Norwo90.02,Wang95.02,vigil98.01,chu05.01}

Recently, using  different methods, a variety of dye-doped polymers were observed to decay and then recover from lasing damage when left to rest.  Using fluorescence measurements, Peng \textit{et al}  reported that rhodamine-doped and pyrromethene-doped polymer optical fibers recovered after photodegradation.\cite{Peng98.01} Howell and Kuzyk \cite{howel02.01,howel04.01} used amplified spontaneous emission(ASE) to observe decay and recovery of 1-amino-2-methylanthraquinone(disperse orange 11, or DO11) dye doped poly(methyl methacrylate) (PMMA).  ASE was found to recover fully even when the damage of the dye-doped polymer was nearly 100$\%$.  However, no recovery was observed in liquid solution of DO11 in MMA - the liquid monomer used to make PMMA polymer.

These results suggest that the host polymer plays an important role in self healing.  While orientational hole burning due to molecular hole burning is a common phenomenon observed in such dye doped polymer systems \cite{abbas-dichroism}, no dichroism is observed at the higher pump intensities used by Embaye and coworkers, which rules out molecular reorientation as the underlying mechanism.\cite{embay08.01}  We also find in our present imaging studies that there is no dichroism associated with the burn track.  Embaye also complemented ASE measurements of DO11-doped PMMA thin films with linear absorption spectroscopy, which showed that during damage, the DO11 molecule is transformed into a quasi stable damaged state, as suggested by an isosbestic point.  Zhu \textit{et al.}\cite{Zhu07.01, Zhu07.02} used two-photon fluorescence of AF455-doped PMMA polymer to measure decay and recovery, and confirmed that this system also recovers fully after photodegradation.

The present work seeks to study the evolution of self-healing of DO11 dye-doped PMMA thin films using a change in the absorbance on the damage track as a proxy for the damaged population.  Since the excitation line has a gaussian profile, such a beam will create a burn mark with a gaussian damage profile. This allows the study of recovery rates as a function of the degree of initial damage by imaging the back-illuminated burn mark of the sample using blue light. From the time dependance of the intensity at each pixel, the recovery rate at the corresponding position of the sample is determined. By exploring recovery as a function of dose, we can investigate whether or not the recovery process is independent of the degree of damage; or, if there is a correlation between the amount of damage and the rate at which the system recovers.  Such information is an important input into the determination of the underlying mechanisms.

\section{Experimental Setup}
\begin{figure}
\centering
\includegraphics{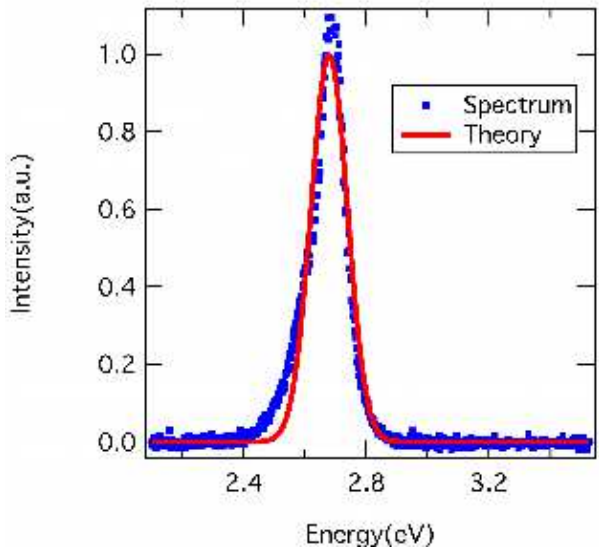}
\caption[LED Intensity]{The LED emission spectrum.  It is sharply peaked($\sigma=14.37 \pm 0.09$ nm) at $\lambda=463$ nm.}
\label{LED}
\end{figure}

Photodegradation and recovery is studied in DO11/PMMA using optical imaging.  The apparatus consists of a computer controlled Nikon Coolpix 990 digital camera with a custom microscope objective mount, a custom sample holder mounted on a three axis translation stage, and a blue LED(Radio Shack 276-0316) peaked at 463nm for controlled sample illumination.  The spectrum is shown in Figure \ref{LED}. An independent UV-Vis absorbance measurement shows that the largest absorption change due to photodegradation is observed in blue part of the spectrum. So, to maximize the observed change in transmittance, we probed the sample with a blue illuminating LED. The LED is connected to a DC power supply via a $97.7 \pm 4.3 k\Omega$ resistor and the applied voltage is maintained at $4.49 \pm 0.01 V$.

The camera was set to use the sRGB IEC61966-2.1 color scheme, with an FN = 4, and an exposure time of 1/60 s.   A 20x objective was attached to the camera and the camera's maximum optical zoom was set to 3x.  By imaging a fixed-width line, it was determined that 1 pixel $ =1.713 \pm 0.062 \mu$m.  For simplicity pixels are used as the unit of length in this paper.

The sample used in this study is a 9g/l thin film of Disperse Orange 11 dye doped in PMMA polymer that is sandwiched between two microscope slides. First, dye is dissolved in MMA and placed in an oven at $95^0C$ for 2 hours to form a highly viscous, semi-polymerized solution. After cooling to room temperature, a few drops of solution are placed on a microscope slide and spread across the slide using another glass slide to form a thin film - a process known as doctor blading. Two such slides were prepared and placed back into the oven for a few more hours. After being fully polymerized, the slides are pressed together - polymer to polymer - and placed in a custom thermal press to form a uniform film. The press temperature is maintained at $150^0C$ and constant uniaxial stress for two hours resulting in uniform thin films. The sample used in the present study is $150 \mu m$ thick.

The thin film is photodegraded with a frequency doubled Nd-YAG laser with the following specifications: 532nm wavelength, 10ns pulse width and operating at 10Hz frequency.  The beam is focused to a line using a cylindrical lens so that it has a Gaussian profile across the line of Gaussian width $\sigma=51.67 \pm 0.03 \mu$m.  The sample was photodegraded by exposing it to the pump beam for 31.4 mins with a per pulse energy of 90$\mu$J.  Details of the experimental setup for burning the sample can be found in the literature.\cite{embay08.01}

\section{Method}
\begin{figure}[h!]
\centering
\includegraphics{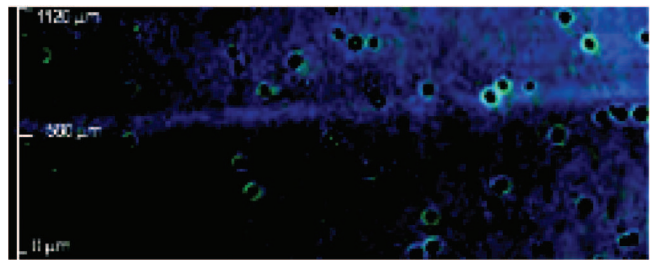}
\caption[Example Burn]{An example of a burn line used in this study(edited for clarity).  Cross sections were taken perpendicular to the burn line with a width of 20 pixels.  The dark spots are the scattering patterns from micro bubbles in the polymer.}
\label{Burn}
\end{figure}

\begin{figure}[h!]
\centering
\includegraphics{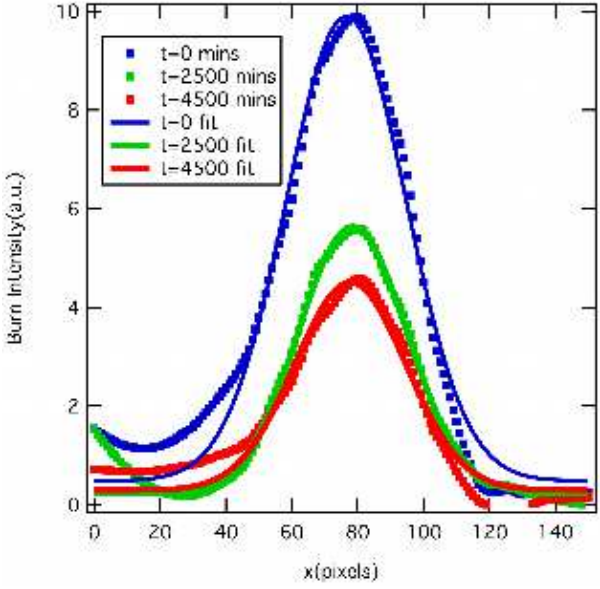}
\caption[Gaussian as a function of x]{Gaussian burn profile of the transmittance as a function of position at several times.  The data has been statistically smoothed for ease of viewing.}
\label{Rawburn}
\end{figure}

After the burn was complete the sample was immediately mounted in the imaging apparatus where it was aligned and brought into focus for the camera.  Images were then taken using exponential time intervals beginning with two-minute intervals, followed by fifteen-, thirty-, and sixty-minute intervals.  The images were then imported into Igor Pro where the RGB line profiles were analyzed at several cross sections with an average burn width of 20 pixels.  Figure \ref{Burn} shows an image of a burn spot.  The burn was found to have a Gaussian profile and was most pronounced in the blue channel, which agrees with the spectral region to be most affected in independent linear absorption measurements.  Using just the blue channel, we fit the Gaussian profile for each image as shown in Figure \ref{Rawburn} for a few representative times. Using the Gaussian fits we determined the value of the new parameter $\beta(x)$, which represents the normalized increase in transmission at position $x$ along the burn, and defined by,
\begin{equation}
\label{eqn:beta}
\beta(x)=\frac{B(x)}{B_0},
\end{equation}
where $B_0$ is the baseline transmitted intensity, and $B(x)$ is the blue channel intensity at position $x$ along the burn.  We define the normalized increase in transmission at the center of the burn as
\begin{equation}
\label{p}
p=\frac{A}{B_0}
\end{equation}
where $A$ is the amplitude of the Gaussian burn profile.

\begin{figure}[h!]
\centering
\includegraphics{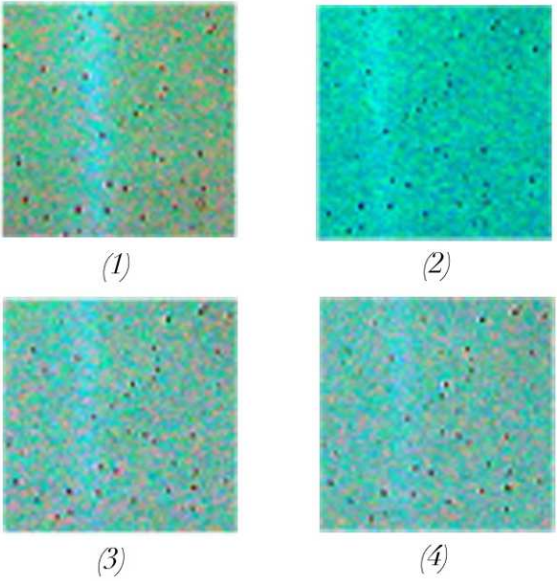}
\caption[Comparison]{Four images of burn lines over time(contrast and color edited for clarity).  1: $t=0$ mins, 2: $t=600$ mins, 3:$t=2739$ mins, 4: $t=5439$ mins. Note the visible dimming of the transmittance profile as a function of time}
\label{Comparison}
\end{figure}

\section{Theory}
In this section, we relate the imaged intensity to the population of damaged molecules.  To determine $\beta$ from the image, it must be recognized that the blue channel intensity, $B$, is a convolution processed in the camera, given by,
\begin{equation}
\label{B(t)}
B(t,x)=\int^{\lambda_{max}}_{\lambda_{min}}S(\lambda)I_0(\lambda)T(\lambda;t,x) d\lambda,
\end{equation}
where $S(\lambda)$ is the sensitivity of the camera's CCD in the range $\lambda_{min}\leq \lambda \leq \lambda_{max}$.  $I_0(\lambda)$ is the LED intensity spectrum and $T(\lambda;t,x)$ is the transmission coefficient of the thin film at time $t$, and position x.

From the Beer-Lambert Law we have,
\begin{equation}
\label{T}
T(\lambda;t,x)=e^{-\sigma(\lambda) n(t,x) l}
\end{equation}
where $\sigma(\lambda)$ is the absorption cross section, $n(t,x)$ is the number density of the absorbers, and $l$ is the thickness of the thin film.

We base our models on observations from photodegradation and recovery studies of ASE in DO11/PMMA that include a long-lived damaged population and an undamaged species that can become damaged through photo-excitation.\cite{embay08.01}  The fractional undamaged population, $n_{u}$, is given by
\begin{equation}
\label{undmg}
n_{u}(t;x)=1-n_0(x)e^{-t/\tau}
\end{equation}
and the fractional damaged population, $n_{d}$, is

\begin{equation}
\label{dmg}
n_{d}(t;x)=n_0(x)e^{-t/\tau}
\end{equation}
where $n_0(x)$ is the fractional initial damaged population at position $x$ along the burn, and $1/\tau$ is the recovery rate.

Recognizing that the damaged and undamaged populations have different absorption cross sections ($\sigma'(\lambda)$ and $\sigma(\lambda)$ respectively) we can substitute Equations \ref{dmg} and \ref{undmg} into Equation \ref{T},

\begin{equation}
T(\lambda;t,x)=e^{-\sigma(\lambda)l\left( 1-n_0(x)e^{-t/\tau} \right)-\sigma'(\lambda)ln_0(x)e^{-t/\tau}},
\end{equation}
and then, rearranging slightly, we get

\begin{equation}
\label{T2}
T(\lambda;t,x)=T_0(\lambda)e^{-\Delta\sigma(\lambda)n_0(x)le^{-t/\tau} },
\end{equation}
where $\Delta\sigma(\lambda)=\sigma'(\lambda)-\sigma(\lambda)$ and $T_0(\lambda)=e^{-\sigma(\lambda)l}$.  Substituting Equation \ref{T2} into Equation \ref{B(t)} we find

\begin{equation}
\label{B2}
B(t;x)=\int^{\lambda_{max}}_{\lambda_{min}}S(\lambda)I_0(\lambda)T_0(\lambda)e^{-\Delta\sigma(\lambda)n_0(x)le^{-t/\tau} } d\lambda.
\end{equation}
If $I_0(\lambda)$ is sufficiently narrow and centered at $\lambda_0$ (as is the case for this experiment) we can approximate it as a delta function and Equation \ref{B2} becomes
\begin{equation}
\label{Bshort}
B(t;x)=S(\lambda_0)I_0(\lambda_0)T_0(\lambda_0)e^{-\Delta\sigma(\lambda_0)n_0(x)le^{-t/\tau} }.
\end{equation}

Substituting Equation \ref{Bshort} into Equation \ref{eqn:beta}, we get the transmittance as a function of position and time,
\begin{equation}
\label{model}
\beta(t;x)=A e^{-\alpha(x) e^{-t/\tau}} ,
\end{equation}
Where $A$ is related to the sensitivity and transmitted intensity, $\alpha(x)=\Delta\sigma(\lambda_0)n_0(x)l$, which is related to the change in absorbance, and $1/\tau$ is the recovery rate.

Assuming that the degree of damage is proportional to the beam intensity, which has a Gaussian profile,  $\alpha(x)$ will be of the form
\begin{equation} \label{profile}
\alpha(x) = \alpha(x_0) e^{-(\frac{x-x_0}{\sigma})^2} ,
\end{equation}
where $x_0$ is position of the center of the burn, $\alpha(x_0)$ the change in transmittance at the burn center,  and $\sigma$ is the beam width.

Because the change in absorption is small relative to the background intensity, the intensity difference determined directly from the raw image is noisy.  The intensity profile for further analysis is determined by the values returned by the fitting function.  Thus, the intensity at each pixel position is determined by first fitting the image data to a Gaussian, and then using the fit parameters  $x_0$, $A$ and $\sigma$ to determine $\beta(t;x)$.

\section{Data and Results}

We first studied the recovery process at the center of the burn, $\beta(t;x_0)$, which is defined as $p$ in Equation \ref{p}.  Using Equation \ref{model}, we performed multiple fits with varying initial parameters and found $\tau=490 \pm 23$ mins.  These results are consistent with literature values of $\tau=476 \pm 14$ mins, which were determined from decay and recovery studies of Amplified Spontaneous Emission(ASE).\cite{embay08.01}  Our only assumption in making this comparison is that the ASE signal is a measure of the undamaged population, or $I_{ASE} \propto n_u(t)$.

\begin{figure}[h!]
\centering
\scalebox{.65}{\includegraphics{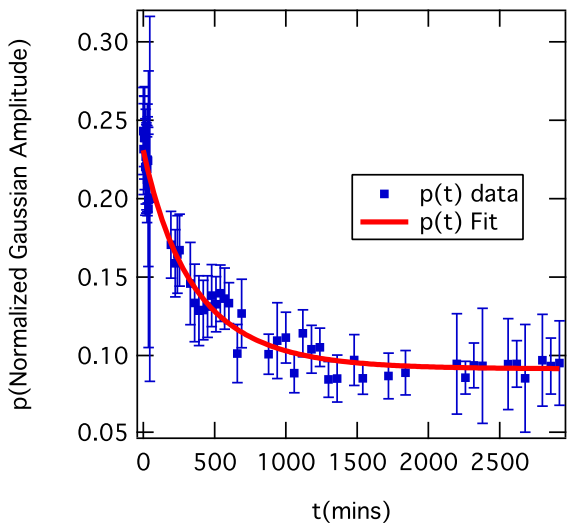}}
\caption[P as a function of time]{The degree of damage as determined from the peak burn dose at the beam center, $p$, as a function of time (points).  The fit (curve) yields $\tau=490 \pm 23$ mins.}
\label{pedit}
\end{figure}

\begin{figure}[h!]
\centering
\includegraphics{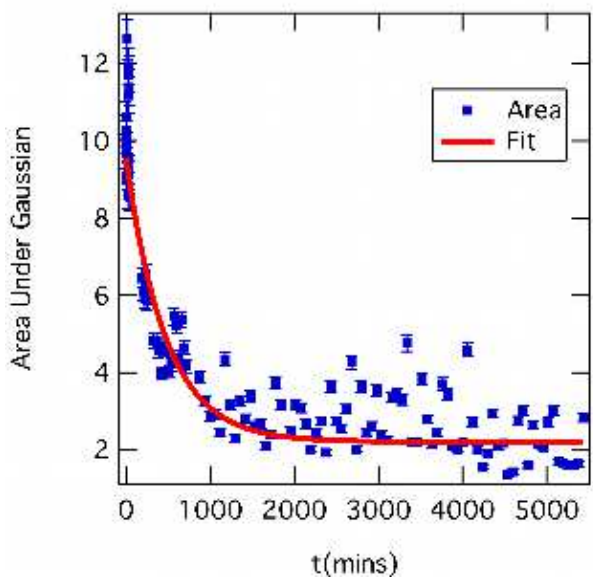}
\caption[Area]{The transmitted intensity averaged over the transverse profile of the Gaussian burn as a function of time (points).  The exponential fit (curve) yields a time constant of $\tau=472.3 \pm 8.8$ mins.}
\label{Area}
\end{figure}

We can explain the small discrepancy in time constants by noting that the ASE intensity, $I_{ASE}$, is the total intensity that originates from all parts of the excited sample.  Thus, the ASE intensity is spatially integrated over the full burn profile, or
\begin{equation}
I_{ASE}(t) \propto \int n_u(t,x) dx .
\end{equation}
If we consider the area under the Gaussian burn as a function of time, as shown in Figure \ref{Area}, and fit it to an exponential (which is the equivalent procedure used for the ASE measurement), we find $\tau=472.3 \pm 8.8$ mins.  This is in agreement with the ASE measurements.

The time constant, $\tau$, which is determined from the time-dependence of the increased transmittance at the center of the burn mark profile, $p$, gives the most precise measurement of the recovery time due to the fact that the change in transmittance there is the largest.

Next we determine the recovery rates as a function of distance from the peak to determine the recovery rates as a function of dose.  Since we postulate that the recovery rate is independent of dose, the time constants should be independent of $\alpha(x)$ since it is proportional to the dose.

\begin{figure}[h!]
\centering
\includegraphics{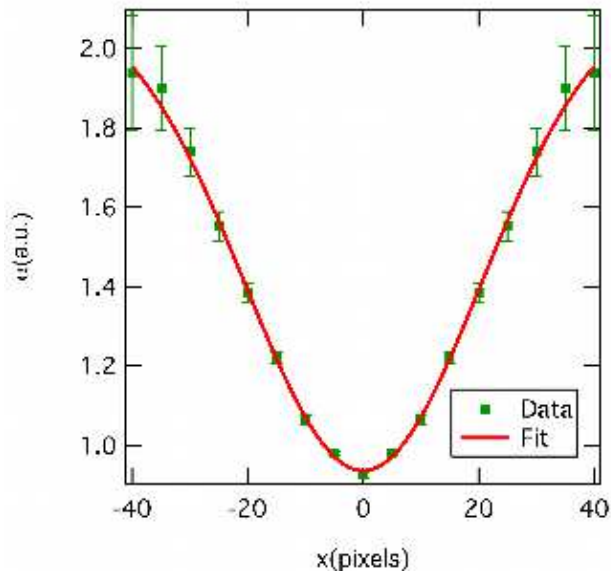}
\caption[Alpha as a function of x]{Using fit parameters we found $\beta(t;x)$, which we then fit to find $\alpha(x)$,which is a Gaussian  across the burn with a width of $\sigma=28.8 \pm 1.9$ pixels, or $\sigma=49.3 \pm 3.7 \mu$m.}
\label{alphax}
\end{figure}

\begin{figure}[h!]
\centering
\includegraphics{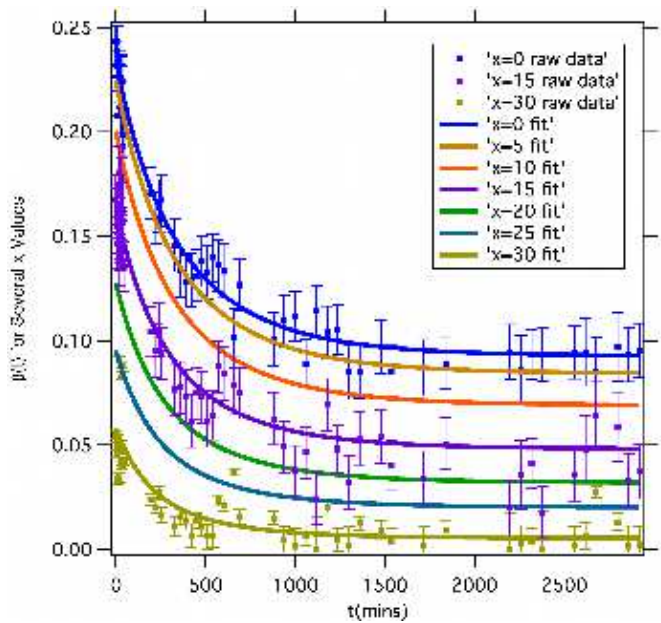}
\caption[Beta as a function of time]{The image intensity as a function of time for several pixels (points) and fits to the function $\beta$ (curves).  To improve plot's clarity, only three raw data sets are shown.  Fitted curves for the omitted data are plotted.}
\label{beta}
\end{figure}

Given that the image is noisy, especially in regions where the dose is small, we use fits to the burn profile to determine the transmittance intensity at each pixel, as follows.  At each time, we fit the data to a Gaussian to get the amplitude and width.  Then using the fitted amplitude and width, we plotted the transmittance at a given $x$ as a function of time, see Figure \ref{beta}.   By fitting this data to our model for the transmittance, Equation \ref{model}, we find that $\alpha(x)$ is indeed Gaussian (see Figure \ref{alphax}) with a width of $\sigma=28.8 \pm 1.9$ pixels or $\sigma=49.3 \pm 3.7 \mu$m, which is within one standard deviation of a direct measurement of beam at the line focus using a beam profiler.  Using the fit parameters, we determine the degree of recovery as a function of the distance from the center.  We can thus determine the degree of population recovery at each point on the burn profile.

Figure \ref{beta} shows the change in transmittance as a function of time for several distances from the center of the burn line as quantified by the pixel number.  The intensity at each pixel is determined using the fit parameters, as described just after Equation \ref{profile}.  The solid curves are the associated fits to the data of the master function $\beta(t;x)$.  Since that data is noisy, only three sets are displayed.  Where data is not shown, the fit functions are shown.  Fitting $\beta(t;x)$ to each of the decay curves determines the recovery time constant. The average time constant from these curves is found to be $\tau= 483 \pm 122$mins, which is within experimental uncertainties of the value found for $p(t)$.  The large standard deviation is due to low signal-to-noise for the lowest intensity curves.

\section{Conclusions}

Using optical imaging studies, we characterized the recovery of a 9g/l DO11/PMMA as a function of time and burn intensity using position from the burn center as a proxy for intensity. Across most of the Gaussian burn, the damaged populations recovered at the same rate, which shows that the rate of recovery is independent of the dose, supporting the argument in our previous work that the recovery process is independent of pump intensity.

Further variables to be considered in future research are the effects of dye concentration, temperature, and polymer composition on the photodegradation and recovery rates.   If the recovery mechanism depends on interactions between the dye molecules, as suggested in previous work,\cite{howel02.01,howel04.01,kuzyk06.06, embay08.01} then the recovery times should depend strongly on concentration.  If the polymer plays a role in recovery, changing the polymer or the temperature (which softens the polymer) should change the recovery rate.

The observation that ASE appears to fully recover over time still needs to be reconciled with the fact that our data suggests that the optically visible damage does not.  Some of the optical imaging data suggests that there is a damage threshold above which a material will not fully recover.  This effect is observed in the tails of the Gaussian burn mark where the burn fully recovers, but near the center, it does not.  The level of noise in our current experiment makes it impossible to determine whether such a full recovery threshold is a property of the material. A more sensitive CCD camera with a higher signal to noise ratio should be able to resolve this problem.

We would like to thank Wright Patterson Air Force Base, Air Force Office of Scientific Research (FA9550-10-1-0286), and NSF (ECCS-0756936) for their continued support of this research.


\end{document}